\documentclass[journal]{IEEEtran}
\usepackage[T1]{fontenc}

\ifCLASSINFOpdf
\else
\fi
\usepackage{amsmath}
\usepackage{makecell}
\usepackage{xspace}
\newcommand*{\eg}{e.g.\@\xspace}
\newcommand*{\ie}{i.e.\@\xspace}

\makeatletter
\newcommand*{\etc}{%
	\@ifnextchar{.}%
	{etc}%
	{etc.\@\xspace}%
}
\usepackage[caption=false]{subfig}
\usepackage{graphicx,amssymb}
\usepackage{url}
\usepackage{hyperref}
\usepackage{amsmath}
\usepackage{multirow}
\usepackage{rotating}
\usepackage{booktabs, multirow} 
\usepackage{changepage,threeparttable} 
\usepackage[table]{xcolor} 
\usepackage{nccmath}
\usepackage{amsmath} 

\newcommand{\comment}[1]{}
\newcommand{\cmnt}[1]{\textcolor{red}{#1}}

\begin{document}\sloppy
%
\title{An Adaptable Deep Learning-Based Intrusion Detection System to Zero-Day Attacks}
%
%
%

\author{Mahdi~Soltani, Behzad~Ousat, Mahdi~Jafari~Siavoshani, Amir~Hossein~Jahangir%
\thanks{M. Soltani, B. Ousat, M. Jafari~Siavoshani, and A. H. Jahangir are with the Department of Computer Engineering, Sharif University of Technology, Tehran, Iran. (email: mahdi@ce.sharif.edu; ousat@ce.sharif.edu; Corresponding author: mjafari@sharif.edu; jahangir@sharif.edu)}}
\maketitle


\begin{abstract}
	The intrusion detection system (IDS) is an essential element of security monitoring in computer networks. An IDS distinguishes the malicious traffic from the benign one and determines the attack types targeting the assets of the organization. The main challenge of an IDS is facing new (\ie, zero-day) attacks and separating them from benign traffic and existing types of attacks. Along with the power of the deep learning-based IDSes in auto-extracting high-level features and its independence from the time-consuming and costly signature extraction process, the mentioned challenge still exists in this new generation of IDSes.
	
	In this paper, we propose a framework for deep learning-based IDSes addressing new attacks. This framework is the first approach using both deep novelty-based classifiers besides the traditional clustering based on the specialized layer of deep structures, in the security scope. Additionally, we introduce DOC++ as a newer version of DOC as a deep novelty-based classifier. We also employ the Deep Intrusion Detection (DID) framework for the preprocessing phase, which improves the ability of deep learning algorithms to detect content-based attacks. We compare four different algorithms (including DOC, DOC++, OpenMax, and AutoSVM) as the novelty classifier of the framework and use both the CIC-IDS2017 and CSE-CIC-IDS2018 datasets for the evaluation. Our results show that DOC++ is the best implementation of the open set recognition module. Besides, the completeness and homogeneity of the clustering and post-training phase prove that this model is good enough for the supervised labeling and updating phase.
\end{abstract}

\section{Introduction}
\label{intro}
An intrusion detection system (IDS) is one of the main layers of defense against cyber threats. Intrusion detection systems have been around for a long time; ranging from the traditional ones, that use the signatures of attacks, to the more recent systems developed using machine learning methods. The process of extracting the signature of attacks is complicated and time-consuming. Moreover, these methods only work for attacks already detected and analyzed, but they are vulnerable to new attacks that have never been detected before.
The newer versions of IDSes are signatureless. These types of IDSes use ML-based methods, especially the deep-learning ones, for the detection of attacks. In \cite{soltani2020content}, the authors have developed a deep learning-based IDS framework with the ability to classify different attacks/benign traffic flows concerning the flow contents.

With all the advances in ML-based IDSes, the signature extraction process and even the feature extraction phase of the suspicious traffic are thoroughly studied. However, another problem still remains: The available IDSes can only detect the previously known attacks. As a result, zero-day attacks remain undetected for a while, and during this period, they can have huge impacts on networks. In \cite{bilge2012before}, the implications of zero-day attacks are discussed. The authors highlight that zero-day attacks are significantly more prevalent than suspected, demonstrating that, out of their 18 analyzed attacks, 11 (61\%) were previously unknown. Having these considerations in mind, the requirement for an IDS which can detect unknown attacks magnifies.

Although there are some statistical and ML-based anomaly detectors in the intrusion detection scope, they are focused on distinguishing the normal traffic from the attacks. As a result, they cannot determine the attack type of the malicious traffic as an applicable detection report. Another weakness of these detectors is the inability to distinguish the new behavior of benign traffic included in the unknowns. Another considerable challenge of traditional ML-based anomaly detectors (based on traditional clustering algorithms) is the high dimensionality of the content of network flows. The high dimensionality of the input is the main reason for using the deep learning-based models in intrusion detectors.

The novelty of this paper is to propose a deep learning-based framework for the adaption of intrusion detectors to zero-day attacks. The framework aims to report the detailed attack-type of the malicious traffic while considering new attacks and new behavior of benign flows. It clusters the unknown samples according to the corresponding new categories. Then, an expert labels these clusters, and the framework updates itself using the new labeled classes. This process enables the framework to update over time.

To the best of our knowledge, this paper uses open set recognition in deep learning-based intrusion classifiers for the first time in the network security scope. Besides, for gathering more evidence of the new attack toward the analysis and updating phases, the traditional clustering algorithm is combined with the deep learning-based classifiers. This combination of clustering with the classification is used for the first time in the intrusion detection scope. Then, the proposed framework is evaluated and compared according to the four models, including DOC \cite{DOCshu2017doc}, DOC++ (which is a newer version of DOC introduced in this paper), Openmax \cite{OpenMaxbendale2016towards}, and a combination of the deep auto-encoder model and traditional SVMs. The evaluation is done over the CIC-IDS2017 \cite{Soheily2017ISCX} and CSE-CIC-IDS2018 \cite{CIC2018} datasets, which are among the most complicated and updated datasets and include the raw content of traffic flows.

The rest of this paper is organized as follows. In the next section, we first review the related works in deep learning-based intrusion detectors, anomaly-based intrusion detectors, and open set recognition in other but similar scopes. In Section \ref{sec:framework}, we describe the proposed framework for intrusion detectors. Section \ref{sec:evaluation} presents details of experiments, dataset preprocessing, and different architectures. The experimental results and comparison of the proposed implementations of the framework are presented in Section \ref{sec:results}. Section \ref{sec:discussion} discusses the best deployment of the framework and the similarities of the labels. Finally, Section \ref{sec:conclusion} concludes the paper.

\section{Related Works}\label{sec:related}
In this section, related works to this project are reviewed in two subsections, namely,
\begin{itemize}
	\item Machine Learning-Based Intrusion Detection Systems
	\item Novelty Based Detectors
\end{itemize}

\subsection{Machine Learning-Based Intrusion Detection Systems}
Machine learning-based intrusion detection systems can be classified into two categories: traditional and deep learning-based approaches.




To summarize, we review some primary research studies on traditional models in ML-based IDSes. The SVM, as the most famous traditional classifier algorithm, is used in researches such as \cite{Heba2010Principle}, \cite{chen2009using}, \cite{jia2017application}, and \cite{wang2017effective}. On the other hand, the $k$-nearest neighbors (KNN) algorithm is popular for unsupervised applications. In \cite{li2014new}, the KNN algorithm is used as the core of an IDS. Random forests (RF) \cite{Breiman2001RF} is a robust algorithm against overfitting with the ability to handle unbalanced data that has been used in works like \cite{Zhang2008RF} and \cite{farnaaz2016random} as the core algorithm of the ML-Based IDS.

In contrast to the traditinal ML models, the complexity of new attacks and the importance of content-based attacks lead to using deep learning-based architectures such as recurrent neural networks, convolutional neural networks, and stacked auto-encoders.

\comment{
\cmnt{Artificial neural networks (ANNs) are one of the most common architectures in deep learning models.} ANNs can be used to detect of DoS attacks like SYNFLOOD, UDPSTORM, and SMURF (for example, see \cite{Palagiri2002ANN}). The authors of \cite{Palagiri2002ANN} use a preprocessing phase with the aid of an anomaly-based ANN, namely, a self-organization map (SOM). This model is evaluated by DARPA 1999 dataset \cite{Lippmannm2000DARPA99} and reaches 100\% detection of normal traffic and 76\% false-positive rate for attacks. 
}

Recurrent neural network (RNN) and long short-term memory (LSTM) network are deep models that use memories to take advantage of the information from previously seen inputs in the judgment of the current one. In particular, LSTM can effectively learn the relations between items far away from each other in a sequence. Computer network flows, consisting of packets, form a data sequence; hence, RNN and LSTM are natural candidates for analyzing computer network traffics. Kim et al. \cite{Kim2016LSTM} have applied LSTM architecture in IDS and use the KDD99 dataset for evaluating their proposed model. Also, authors in \cite{Agarap2018GRU} have employed gated recurrent unit (GRU), a variant of LSTM network. They have evaluated the proposed model with network traffic data captured by the honeypot systems at Kyoto University in 2013.



Convolutional neural network (CNN) is another architecture used in deep learning models that take advantage of shared weights in different sliding windows and are space invariant. Research studies like \cite{li2020robust}, \cite{kwon2018empirical}, and \cite{kim2020cnn} use CNN for building an ML-Based IDS. In \cite{kim2020cnn}, CNNs are used to detect DoS (denial of service) attacks of two datasets: KDD99 (as a typical dataset) and CSE-CIC-IDS2018 (as the most up-to-date IDS dataset).




In some other research studies such as \cite{Aminanto2018SAE}, \cite{Javaid2016AE}, \cite{manzoor2017feature}, \cite{aminanto2017deep}, \cite{javaid2016deep}, and \cite{Aminanto2017SAE}, the deep learning approach is employed for the reduction of input dimensions by selecting among pre-extracted features.

\subsection{Novelty-Based Detectors}
Zero-day attacks are one of the major concerns in intrusion detection systems. These attacks constitute the main weakness of the traditional signature-based IDSes. Since signatures are extracted based on the known attacks, the traditional approach is vulnerable to zero-day attacks that happen for the first time. This issue is also present in the learning-based IDSes. In learning-based models, zero-day attack detection can be considered novelty detection in open set recognition.

Learning-based detectors can be categorized into two main groups: classification-based and anomaly-based learning. Anomaly-based learning models \cite{aldweesh2020deep} are capable of detecting traffic that is abnormal. However, their main weakness is distinguishing and reporting the detected attacks. On the other hand, classification-based models can report the sub-category of the known attacks, but they are vulnerable to zero-day attacks just like the signature-based detectors. This paper aims to cover the weaknesses of the above-mentioned learning-based models. It reports the category of known and unknown (\ie, zero-day) attacks simultaneously.

Studies around anomaly-based detectors besides novelty detection algorithms in non-security scopes are reviewed in the following.
In \cite{naseer2018enhanced}, the authors investigate the suitability of deep learning models (such as CNN, auto-encoder (AE), and RNN) for anomaly-based detection systems.
Although studies such as \cite{kwon2018empirical}, \cite{naseer2018enhanced}, and some other ones mentioned in \cite{aldweesh2020deep}, are labeled as anomaly-based intrusion detectors, they mainly consider the attack traffic as abnormal (even though this attack may have been seen in the training phase). In contrast, we focus on the detection of the new attacks which have not been seen in the training phase.


In \cite{aminanto2017improving}, a combination of stacked auto-encoder (SAE) and clustering is implemented. The SAE is responsible for auto feature extraction and these features are used as the input of the $k$-means clustering algorithm with two clusters (\ie, benign and anomaly traffic). This clustering algorithm is the key solution for distinguishing the novelty attack in Wi-Fi networks.



Authors of \cite{alom2017network} detect anomaly attacks using iterative $k$-means clustering. In \cite{yang2017towards}, dimensionality reduction problem is combined with clustering to enable the deep model to extract suitable features for clustering.


Novelty detection in the literature is also known as \emph{open set recognition}. In \cite{geng2020recent}, the authors have done survey research on open set recognition methods and classified them as traditional ML methods, DNN based methods, and generative models.

In \cite{OpenMaxbendale2016towards}, Bendale and Boult have proposed a solution called OpenMax for open set recognition. In this method, the model is trained with the normal Softmax cross-entropy loss, then in the test phase, each class is represented as a mean activation vector (MAV) which is calculated as the mean of the values in the penultimate layer of the network for correctly classified training samples. Then, the training samples' distances from their corresponding MAVs are calculated and used to fit a Weibull distribution for each known class. The final score for classification is computed using Softmax on these distributions. 

By combining the ideas of OpenMax and encoders, another method is introduced in \cite{CROSR}, called classification-reconstruction learning for open set recognition (CROSR). In this method, instead of using the penultimate layer to calculate the activation vector, they use the latent representation extracted from their design of deep hierarchical reconstruction nets (DHRNets) which provide a prediction alongside the latent representation for each sample.

The other research inspired by the OpenMax method is described in \cite{DOCshu2017doc}. In this research, Shu et al. have proposed a new method called deep open classification (DOC). They have replaced the Softmax loss layer of the model with a 1-vs-rest layer consisted of sigmoids. In the test phase, the method simply applies a threshold on the maximum output value to determine the class of the sample being tested and classifies it as one of the known classes or rejects as unknown.

\section{Framework}\label{sec:framework}
In this section, the proposed novelty-based framework for deep learning-based intrusion detectors is described.  As applicable intrusion detectors should cover the wide range of new attacks that are emerging continuously, this framework has to adapt the deep classification models with zero-day attacks in the real world's circumstances. This framework consists of four phases.
\begin{itemize}
	\item The first phase distinguishes the new attacks from the older ones. The main weakness of the known anomaly-based detectors is that they just label the anomaly inputs. However, in the real world devices, the detailed report of the security incidents, is the primary concern. Besides, anomaly traffic is not essentially malicious. It may be generated by a new behavior of benign users or just be a new form of the known types of attacks.
	\item The second phase gathers enough evidence and instances of the same new attack for deeper analysis by security experts.
	\item In the third phase, the expert supervisor determines the types of unknown traffic in one of these four subcategories: known malicious traffic, new attack, unseen benign, and temporary anomaly traffic.
	\item Finally, in the fourth phase, the deep model will be updated using the new information.
\end{itemize}

Open set recognition is the main contribution of the first phase of the framework. Additionally, as the deep learning models need enough instances of a new attack for the update phase, the evidence gathering for these types of attacks is implemented by merging the traditional clustering algorithms by the nature of the deep learning models. To the best of our knowledge, the concept of open set recognition and its combination with clustering has been applied for the first time in the context of intrusion detection. In the following, the details of the phases of the proposed framework are discussed.

\begin{figure}
	\centering
	\includegraphics[width=0.78\linewidth]{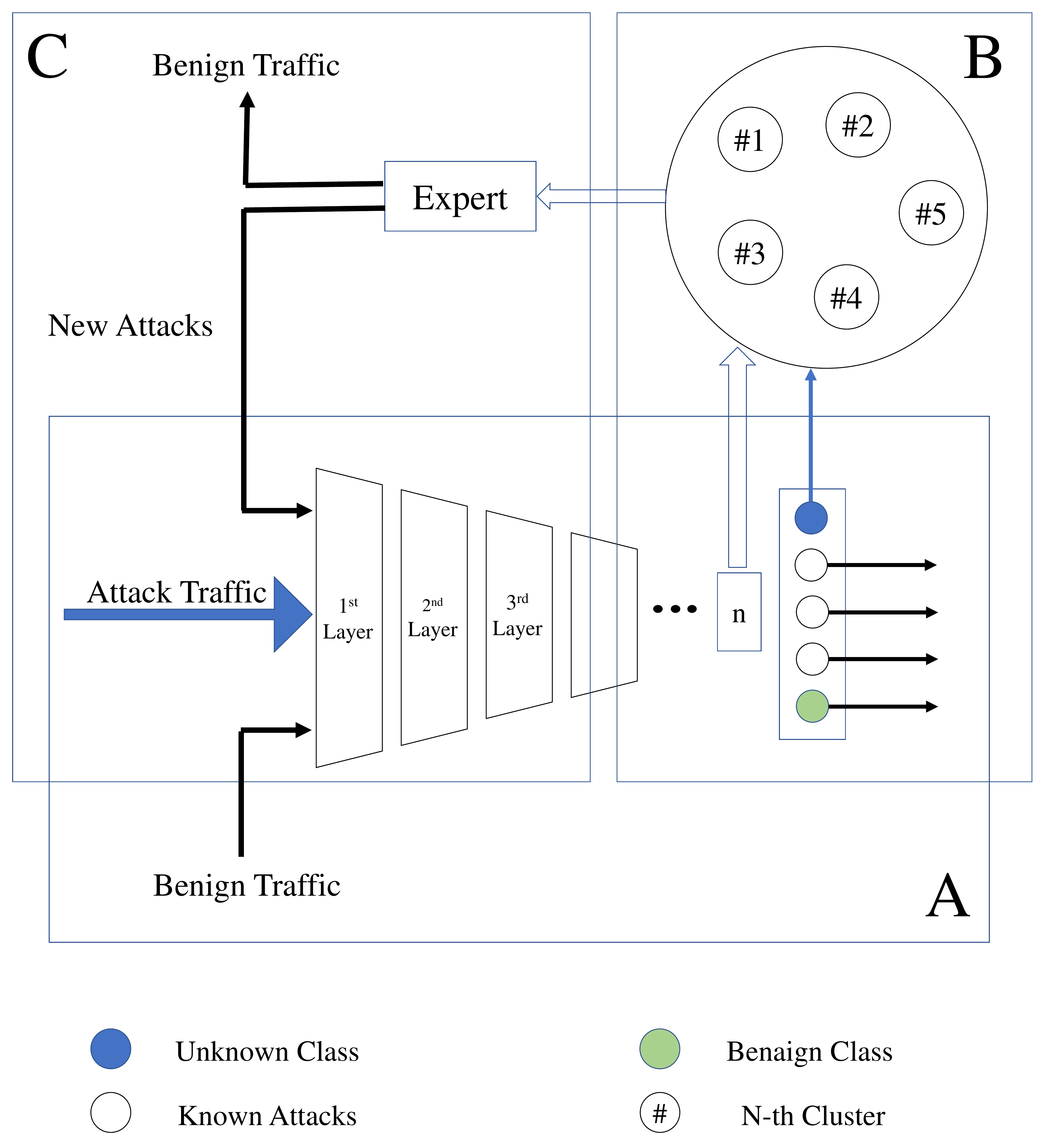}
	\caption{High level framework modules: (A) Open set recognition, (B) Clustering, and (C) Supervised labeling and updating the model.}
	\label{fig:framework}       
\end{figure}

\subsection{Phase I: Open Set Recognition Module}
The open set recognition module is responsible for detecting unknown traffic. This module has a prominent role in the framework for determining the unknown traffic. In this paper, we evaluate the following methods for the implementation of the proposed framework:

\subsubsection{Deep Open Classification} \label{sub: doc}
To apply DOC for our framework, we need to train a model using the 1-vs-rest loss function \cite{DOCshu2017doc}, as follows
\begin{multline}
\label{1-vs-rest loss}
L = \sum_{i=1}^{m}{\sum_{j=1}^{n}{-I(y_{i} = l_{i}) \log{p(y_{j} = l_{j})}}}\\
- I(y_{i} \neq l_{i}) \log{(1 - p(y_{j} = l_{j}))}.
\end{multline}
 The loss layer contains $m$ sigmoid functions for $m$ known classes. In \eqref{1-vs-rest loss}, $I$ is the indicator function and $p(y_{j} = l_{i})$ is the sigmoid output of the $i$-th neuron for $j$-th sample.
The equation consists of two parts. The first one optimizes the correct classification, and the second one reduces the misclassifications of class $l_{i}$.
In the evaluation phase, it uses a threshold to determine the class of each sample or reject it based on \cite[Equation~3]{DOCshu2017doc}.

\subsubsection{DOC++} \label{sub: doc++} 
The main weakness of DOC is that the model tries to classify all input classes in the training phase. As a result, it cannot learn the assumptions about novelty classes that do not belong to the known classes. This learning phase is not compatible with our expectations in the testing phase. According to our evaluations, DOC can perform better by some modifications.
DOC++ is introduced in this paper for making a better implementation of the deep learning-based novelty detector IDS.

The core idea of DOC++ is to instruct the model about the existence of unknown classes. 
Consequently, we provide some extra unknown labels as the input in the training phase. In this case, the first part of the 1-vs-rest loss function stated in \eqref{1-vs-rest loss}, which corresponds to correct classification, equals zero, and the second part tries to minimize the classification probability of the unknown labels as known classes. As a result, the second part of the equation minimizes the corresponding loss value of these unknown labels. The evaluation phase of DOC++ is the same as DOC. However, the unknown labels used in the training phase are eliminated from the evaluation.

\subsubsection{OpenMax} \label{sub: openmax}
As mentioned previously, OpenMax relies on MAVs which are calculated based on \cite[Algorithm~1]{OpenMaxbendale2016towards}. Next, the Weibull distributions and the new scoring vectors are computed based on \cite[Algorithm~2]{OpenMaxbendale2016towards}.
Note that the training phase utilizes a simple Softmax classification loss.
\subsubsection{AutoSVM} \label{sub: autosvm}
The last implementation of the open set recognition module is the combination of one-class SVM and stacked auto-encoder (SAE). The idea is to use a one-class SVM for each known class to decide whether a sample belongs to the SVM's corresponding distribution or not. Finally, the samples that are rejected by all of the SVMs are labeled as unknown traffic. The problem with this naive method is that SVMs do not perform well on samples with high dimensionality. To solve this issue, an auto-encoder is used for the dimension reduction phase. Hence, high-level essential features are extracted from the auto-encoder and are fed to the SVM models.

\subsection{Phase II: Clustering and Post Training}
We need to gather enough data from the unknown distributions and retrain the model to keep the model updated.
Besides, the multi-instances of new classes of traffic may be observed in a specific time window. This phase is responsible for separating the instances of the corresponding different novel classes while providing enough evidence of each class as the inputs of the following modules of the framework.

This phase is implemented by a clustering module that connects to a specific layer of the deep classifier model and generates clusters of the instances of the observed unknown traffic. In an ideal scenario, each cluster contains just the samples of the same label, and all samples of a label are available only in one cluster.

To further improve the clustering accuracy, we have set up a post-train phase to optimize the deep model to generate clustering-friendly outputs.
To define a post-training loss function we need to label every generated cluster in the training phase. Therefore, each cluster's centroid is assigned a label.
The new loss function $L^{\prime}$ is based on the distance between the output of the \emph{particularized layer} $\hat{x}_{i}$ and the same-label centroid $C_{y_{i}}$, as follows
\begin{equation}\label{cluster-loss}
	\begin{gathered}
		L^{\prime} = \sum_{i=1}^{n}\lVert {\hat{x}_{i}} - C_{y_{i}} \rVert^2 ,\\ 
		C_{y_{i}} = \text{Centeroid with the same label as $y_{i}$,}
	\end{gathered}
\end{equation}
where $x_i$ is the input and $y_i$ is its label.
The particularized layer used in the clustering is different in each architecture:
\begin{itemize}
	\item In DOC, the penultimate layer is used for the clustering and post-train process,
	\item In OpenMax, the clustering is applied to the recalibrated scores after removing the unknown score,
	\item In AutoSVM, the output of the encoder part of SAE is used as the input to the clustering module.
\end{itemize}


\subsection{Phase III: Supervised Labeling}
Once a sufficient number of instances is provided for some specific unknown class, we need to retrain the model with all new known classes (\ie, at least one class is added to the pre-known labels). The label of the new classes will be determined by an expert, according to the gathered evidence. This label can be on the three main categories:
\begin{itemize}
	\item Malicious traffic: The expert will label the new classes that belong to this type of traffic by the name of the new kind of attack. The instances of this attack will be added as a new class in the retraining phase of the deep model.
	\item Unseen benign traffic: The novelty traffic may be unseen benign traffic. Consequently, the instances of this novelty class will be merged with the existing normal instances for the retraining phase.
	\item Temporary anomaly traffic: In some conditions, like the university semester registration time, anomaly traffic is neither malicious nor unseen normal traffic that we expect to see in the future. These labeled classes are ignored in the retraining phase.
\end{itemize}

\subsection{Phase IV: Updating the Model}
To keep the framework up and running in the retraining phase, we propose an active-passive strategy as the following steps.
\begin{enumerate}
	\item Clone the existing active model to a passive model.
	\item Run the training, clustering and, post-training phases on the cloned model.
	\item Migrate the traffic to the new model.
\end{enumerate}

\section{Experimental Evaluation}\label{sec:evaluation}
\label{experiment}

This section describes the details of the evaluation of the proposed framework to reproduce the experiments. First, the preprocessing phase of the training and test state is described. Then, the datasets and the reasons for choosing the selected datasets are discussed. Finally, the different implementations of the proposed framework are described\footnote{The implementations of all evaluated models are available at \url{https://github.com/INL-Laboratory/Adaptable-IDS}.}. 

We should mention that the framework's main phases to be evaluated are the first and second phases. Subsequent phases are more descriptive and do not require implementation evaluation. The third phase demands an expert to label the gathered pieces of evidence of the novelty class. Finally, the fourth phase is just retraining the model with the combination of new and previous known attacks. In other words, it is equivalent to just increasing the number of known attacks in our evaluations (for example by changing the number of known classes from four to five in our experiments).

\subsection{Preprocessing}
The Deep Intrusion Detection (DID) approach \cite{soltani2020content} is used in the preprocessing phase of all implemented deep models in this paper. The DID approach is selected for its self-extraction of the appropriate features, besides its capability of detecting a wide range of attacks, including content-based ones like SQL injection and Heartbleed attack. The content-based attacks are the main segment of the threats with high malicious impacts on the targeted organizations. Consequently, this preprocessing phase can have a significant impact on the applicability of the proposed framework. We used a packet size of 200 bytes and a flow size of 100 packets that result in a 20000-dimensional input vector. This selection is based on the analysis of the correspondent datasets, investigated in \cite{soltani2020content}. 

In the evaluations of this work, we need to consider some types of classes as known traffic and the others as the novelty ones. Therefore, we created a pool of all available labels. In the training phase, we select a subset of this pool, and in the test phase, another label is chosen from the pool as an unknown traffic.

\subsection{Dataset}
As the DID approach is designed for the applicable IDSes, it uses the pure content of traffic flows (\eg, in PCAP format). Consequently, the chosen dataset is limited to the datasets that include labeled content of the traffic. The privacy issues restrict the dataset developers from publishing the details of the real network traffic. As a result, datasets with full fraffic content such as DARPA 1999 \cite{lippmann20001999} (which is the base of the KDD99 \cite{KDD99} and NSL-KDD \cite{tavallaee2009detailed} dataset), CIC-IDS2017 \cite{Sharafaldin2018ISCX}, and CSE-CIC-IDS2018 \cite{CIC2018} are all generated in an emulated network.

The main weakness of the emulated datasets is their simplicity: lack of adequate variety in the protocol types and existence of a few numbers of values in different fields in the selected protocols \cite{ring2019survey}. In \cite{Mahoney2003DARPA}, authors have shown that many parameters of the DARPA 99 \cite{Lippmannm2000DARPA99} traffic, like TTL (Time To Live), ToS (Type of Service), and the IP addresses, can cause overfitting. For example, TTLs of the attack traffics are mostly 126 and 253, but benign traffic has nine restricted values, different from the attack ones. Besides, source IP addresses of attacks are different from benign traffics and can simply be used for discrimination. The KDD99 dataset also has inherited these vulnerabilities. Since the attack traffic constitutes a small part of the dataset, there are many purely benign IP addresses. The algorithm can assign a substantial weight for IP addresses to attain higher accuracy. However, we know that this is not a valid assumption in the real world. 

Having all of the mentioned shortcomings in mind, in this work, we have used the more up-to-date datasets (CIC-IDS2017 and CSE-CIC-IDS2018) for the proper evaluation of the framework, which have implemented the more recent attack types like SSH brute force, botnet, DoS, DDoS, web, and infiltration attacks. Most importantly, some content-based attacks like SQL injection, XSS attacks, and Heartbleed exists among them. Additionally, benign profiles are extracted based on the abstract behavior of 25 users based on the HTTP, HTTPS, FTP, SSH, and email protocols.

\subsection{Models' Architecture}
In our experiments, we have implemented four different models as the open set classifier: DOC, DOC++, OpenMax, and AutoSVM. For DOC, DOC++, and OpenMax, the deep neural network is inspired by the model presented in \cite{DOCshu2017doc}. The first two layers consist of a 1-dimensional convolutional layer and a max-pooling layer to extract the best features. Filter size of 20 and an output channel size of 10 are set for the convolutional layer. To further reduce the size of the output vector, we applied two additional fully connected (FC) layers with ReLU6 activation. These layers have 500 and 4 neurons, respectively. Finally, a different loss layer is applied in each model. 

In DOC and DOC++, the output of the last FC layer (the penultimate layer) is considered as the particularized layer for merging with the clustering model. However, the particularized layer of the OpenMax is the recalibrated vector of the calculated Weibull distribution. It is possible to use recurrent neural networks instead of the convolutional layer. Our experiments show an insignificant change in the results with this replacement, especially for the unknown classification (see CNN results in Tables \ref{tab: avg_known_acc} and \ref{tab: avg_unkown_classification} and LSTM results in Tables~\ref{tab: LSTM Known} and \ref{tab: LSTM Unknown}). Therefore, we used the CNN architecture as the original DOC paper \cite{DOCshu2017doc}.

For DOC and DOC++, the novelty detection method described in section \ref{sub: doc} is used. In the OpenMax model, the applied novelty detection method is based on section \ref{sub: openmax}.

The difference between DOC and DOC++ is the number of labeled classes in the training phase. In DOC, we use four different classes in the training phase besides adding a new label as the novelty in the test phase. But in DOC++, the training is done using five labels while keeping the model architecture the same as before (\ie, four output neurons).

The other implemented model is the AutoSVM approach. This model consists of an auto-encoder with an input size of 20000 and a mid-layer size of 1000. Later, the encoded vectors are feed into four different one-class SVMs with parameter $\nu$ equal to 0.01 and radial basis function (RBF) kernel with automatic gamma parameter.

\section{Evaluation Results}\label{sec:results}
In this section, the evaluation results of different implementations of the proposed framework are illustrated. Firstly, we report the results of each framework implementation in a separate subsection. Multiple implementations of the open set recognition modules such as DOC++, DOC, OpenMax, and AutoSVM are evaluated in these subsections. All of these experiments have been done over the CIC-IDS2017 dataset. 

We evaluate two approaches in the evaluation phase. The first one tries to detect the best algorithm for classifying the existing classes, including benign and attack types. The second approach considers that an anomaly detection module separates the normal and anomaly traffic and only forwards the abnormal traffic to the open set classifier. In this approach, benign traffic is absent in the training and test phases. The final goal is to report the known attacks and introduce new ones. The evaluation of this approach is marked as follows: DOC++\_NN, DOC\_NN, OpenMax\_NN, and AutoSVM\_NN.

Finally, the results of the selected model are reported over the CSE-CIC-IDS2018 dataset. This double-checking evaluation is just for making the results more reliable.

\subsection{Classification of Known Labels}
The first experiment is to evaluate the classification accuracy of the model on known labels.
This experiment is a classification in a closed set environment, which the results are provided in Table \ref{tab: avg_known_acc}.
This table shows the average accuracy of the classification per each label. This result is derived from the average results of several different experiments where the specified label is a part of the training phase.

\begin{table}[!htp]\centering
	\caption{Average Accuracy on Closed Set Using CNN Cells.}\label{tab: avg_known_acc}
	\normalsize
	\centering
	\resizebox{\linewidth}{!}{%
		\begin{tabular}{lccccccc}
			\hline
			\parbox[c][0.5cm][c]{2.5cm}{\raggedright Labels} &DOC++ &DOC++\_NN &DOC &DOC\_NN &OpenMax &OpenMax\_NN &AutoSVM\_NN \\
			\hline
			\parbox[c][0.5cm][c]{3cm}{\raggedright \cellcolor[HTML]{b6d7a8}Botnet} &\cellcolor[HTML]{b6d7a8}91.53 &\cellcolor[HTML]{b6d7a8}88.93 &\cellcolor[HTML]{b6d7a8}92.48 &\cellcolor[HTML]{b6d7a8}96.14 &\cellcolor[HTML]{b6d7a8}93.77 &\cellcolor[HTML]{b6d7a8}96.59 &\cellcolor[HTML]{b6d7a8}76.65 \\
			\parbox[c][0.5cm][c]{3cm}{\raggedright \cellcolor[HTML]{b6d7a8}DDoS} &\cellcolor[HTML]{b6d7a8}91.72 &\cellcolor[HTML]{b6d7a8}90.9 &\cellcolor[HTML]{b6d7a8}95.62 &\cellcolor[HTML]{b6d7a8}95.78 &\cellcolor[HTML]{b6d7a8}95.35 &\cellcolor[HTML]{b6d7a8}96.13 &\cellcolor[HTML]{b6d7a8}82.48 \\
			\parbox[c][0.5cm][c]{3cm}{\raggedright \cellcolor[HTML]{ea9999}Port Scan} &\cellcolor[HTML]{ea9999}91.69 &\cellcolor[HTML]{ea9999}91.93 &\cellcolor[HTML]{ea9999}94.44 &\cellcolor[HTML]{ea9999}95.99 &\cellcolor[HTML]{ea9999}95.47 &\cellcolor[HTML]{ea9999}95.78 &\cellcolor[HTML]{ea9999}87.48 \\
			\parbox[c][0.5cm][c]{3cm}{\raggedright Benign Wednesday} &86.39 &N/A &87.04 &N/A &89.97 &N/A &N/A \\
			\parbox[c][0.5cm][c]{3cm}{\raggedright \cellcolor[HTML]{b6d7a8}DoS GoldenEye} &\cellcolor[HTML]{b6d7a8}90.47 &\cellcolor[HTML]{b6d7a8}93.72 &\cellcolor[HTML]{b6d7a8}93.36 &\cellcolor[HTML]{b6d7a8}95.26 &\cellcolor[HTML]{b6d7a8}93.99 &\cellcolor[HTML]{b6d7a8}94.62 &\cellcolor[HTML]{b6d7a8}75.84 \\
			\parbox[c][0.5cm][c]{3cm}{\raggedright \cellcolor[HTML]{b6d7a8}DoS Hulk} &\cellcolor[HTML]{b6d7a8}92.3 &\cellcolor[HTML]{b6d7a8}93.19 &\cellcolor[HTML]{b6d7a8}94.13 &\cellcolor[HTML]{b6d7a8}95.41 &\cellcolor[HTML]{b6d7a8}94.34 &\cellcolor[HTML]{b6d7a8}97.38 &\cellcolor[HTML]{b6d7a8}76.7 \\
			\parbox[c][0.5cm][c]{3cm}{\raggedright \cellcolor[HTML]{a2c4c9}DoS SlowHttp} &\cellcolor[HTML]{a2c4c9}91.75 &\cellcolor[HTML]{a2c4c9}89.76 &\cellcolor[HTML]{a2c4c9}93.5 &\cellcolor[HTML]{a2c4c9}96.1 &\cellcolor[HTML]{a2c4c9}94.67 &\cellcolor[HTML]{a2c4c9}95.89 &\cellcolor[HTML]{a2c4c9}79.98 \\
			\parbox[c][0.5cm][c]{3cm}{\raggedright \cellcolor[HTML]{a2c4c9}DoS SlowLoris} &\cellcolor[HTML]{a2c4c9}91.6 &\cellcolor[HTML]{a2c4c9}90.78 &\cellcolor[HTML]{a2c4c9}94.94 &\cellcolor[HTML]{a2c4c9}96.18 &\cellcolor[HTML]{a2c4c9}94.47 &\cellcolor[HTML]{a2c4c9}94.8 &\cellcolor[HTML]{a2c4c9}78.88 \\
			\parbox[c][0.5cm][c]{3cm}{\raggedright \cellcolor[HTML]{a2c4c9}FTP-Patator} &\cellcolor[HTML]{a2c4c9}92.36 &\cellcolor[HTML]{a2c4c9}92.45 &\cellcolor[HTML]{a2c4c9}94.48 &\cellcolor[HTML]{a2c4c9}95.65 &\cellcolor[HTML]{a2c4c9}95.56 &\cellcolor[HTML]{a2c4c9}95.92 &\cellcolor[HTML]{a2c4c9}84.97 \\
			\parbox[c][0.5cm][c]{3cm}{\raggedright \cellcolor[HTML]{b6d7a8}SSH-Patator} &\cellcolor[HTML]{b6d7a8}92.37 &\cellcolor[HTML]{b6d7a8}93.02 &\cellcolor[HTML]{b6d7a8}94.83 &\cellcolor[HTML]{b6d7a8}95.2 &\cellcolor[HTML]{b6d7a8}94.85 &\cellcolor[HTML]{b6d7a8}96.81 &\cellcolor[HTML]{b6d7a8}72.27 \\
			\parbox[c][0.5cm][c]{3cm}{\raggedright Benign Friday} &88.31 &N/A &88.82 &N/A &90.05 &N/A &N/A \\
			\parbox[c][0.5cm][c]{3cm}{\raggedright \cellcolor[HTML]{ffe599}Web BruteForce} &\cellcolor[HTML]{ffe599}80.91 &\cellcolor[HTML]{ffe599}80.72 &\cellcolor[HTML]{ffe599}84.06 &\cellcolor[HTML]{ffe599}85.57 &\cellcolor[HTML]{ffe599}84.86 &\cellcolor[HTML]{ffe599}87.77 &\cellcolor[HTML]{ffe599}70.09 \\
			\parbox[c][0.5cm][c]{3cm}{\raggedright \cellcolor[HTML]{ffe599}Web XSS} &\cellcolor[HTML]{ffe599}82.2 &\cellcolor[HTML]{ffe599}82.47 &\cellcolor[HTML]{ffe599}85.78 &\cellcolor[HTML]{ffe599}84.38 &\cellcolor[HTML]{ffe599}83.72 &\cellcolor[HTML]{ffe599}86.34 &\cellcolor[HTML]{ffe599}69.03 \\
			\hline
		\end{tabular}
	}
\end{table}

\begin{table}[!htp]\centering
	\caption{Average Accuracy on Closed Set Using LSTM Cells.}
	\label{tab: LSTM Known}
	\scriptsize
	\begin{tabular}{lccc}
		\hline
		\parbox[c][0.25cm][c]{2.5cm}{\raggedright Labels}  & DOC++ & DOC++\_NN \\
		\hline
		\parbox[c][0.25cm][c]{3cm}{\raggedright \cellcolor[HTML]{b6d7a8}Botnet} &\cellcolor[HTML]{b6d7a8}94.8 &\cellcolor[HTML]{b6d7a8}96.03 \\
		\parbox[c][0.25cm][c]{3cm}{\raggedright \cellcolor[HTML]{b6d7a8}DDoS} &\cellcolor[HTML]{b6d7a8}94.22 &\cellcolor[HTML]{b6d7a8}96.11 \\
		\parbox[c][0.25cm][c]{3cm}{\raggedright \cellcolor[HTML]{ea9999}Port Scan} &\cellcolor[HTML]{ea9999}95.03 &\cellcolor[HTML]{ea9999}94.33 \\
		\parbox[c][0.25cm][c]{3cm}{\raggedright Benign Wednesday}  &87.87 & N/A \\
		\parbox[c][0.25cm][c]{3cm}{\raggedright \cellcolor[HTML]{b6d7a8}DoS GoldenEye} &\cellcolor[HTML]{b6d7a8}94.63 &\cellcolor[HTML]{b6d7a8}93.73 \\
		\parbox[c][0.25cm][c]{3cm}{\raggedright \cellcolor[HTML]{b6d7a8}DoS Hulk} &\cellcolor[HTML]{b6d7a8}94.06 &\cellcolor[HTML]{b6d7a8}94.86 \\
		\parbox[c][0.25cm][c]{3cm}{\raggedright \cellcolor[HTML]{a2c4c9}DoS SlowHttp} &\cellcolor[HTML]{a2c4c9}95.98 &\cellcolor[HTML]{a2c4c9}94.81 \\
		\parbox[c][0.25cm][c]{3cm}{\raggedright \cellcolor[HTML]{a2c4c9}DoS SlowLoris} &\cellcolor[HTML]{a2c4c9}94.73 &\cellcolor[HTML]{a2c4c9}93.39 \\
		\parbox[c][0.25cm][c]{3cm}{\raggedright \cellcolor[HTML]{a2c4c9}FTP-Patator} &\cellcolor[HTML]{a2c4c9}95.72 &\cellcolor[HTML]{a2c4c9}92.92 \\
		\parbox[c][0.25cm][c]{3cm}{\raggedright \cellcolor[HTML]{b6d7a8}SSH-Patator} &\cellcolor[HTML]{b6d7a8}94.95 &\cellcolor[HTML]{b6d7a8}94.42 \\
		\parbox[c][0.25cm][c]{3cm}{\raggedright Benign Friday} &89.11 &N/A \\
		\parbox[c][0.25cm][c]{3cm}{\raggedright \cellcolor[HTML]{ffe599}Web BruteForce} &\cellcolor[HTML]{ffe599}84.74 &\cellcolor[HTML]{ffe599}85.03 \\
		\parbox[c][0.25cm][c]{3cm}{\raggedright \cellcolor[HTML]{ffe599}Web XSS} &\cellcolor[HTML]{ffe599}85 &\cellcolor[HTML]{ffe599}82.12 \\
		\hline
	\end{tabular}
\end{table}

\subsection{Classification of Unknown Labels} \label{sub: unkown_classification}
\begin{figure}
	\centering
	\scriptsize
	\begin{tabular}{cc}
		\includegraphics[width=40mm]{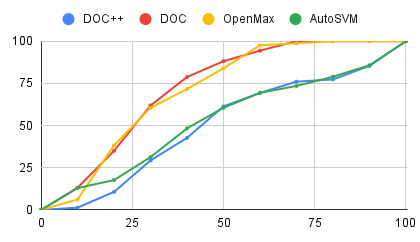} &   \includegraphics[width=40mm]{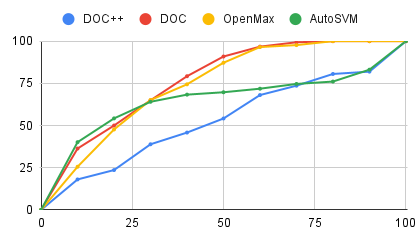} \\
		DoS GoldenEye & DoS Hulk \\[6pt]
		\includegraphics[width=40mm]{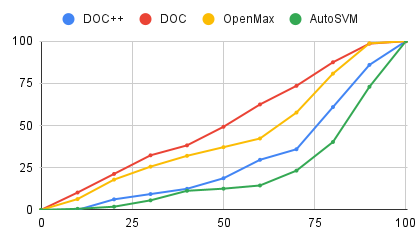} &   \includegraphics[width=40mm]{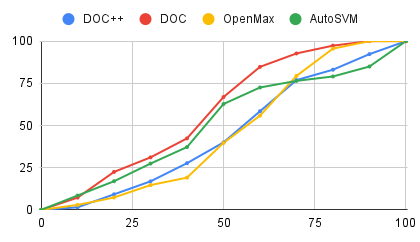} \\
		DoS SlowHttp & DoS SlowLoris \\[6pt]
		\includegraphics[width=40mm]{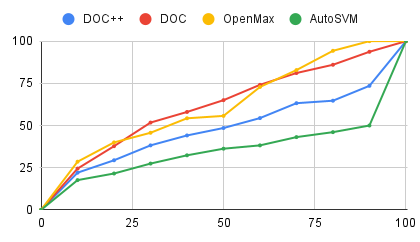} &   \includegraphics[width=40mm]{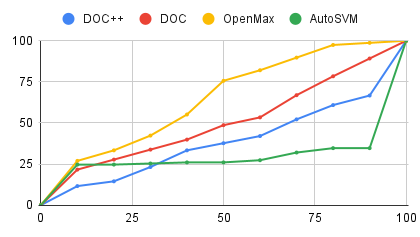} \\
		FTP-Patator & SSH-Patator \\[6pt]
		\includegraphics[width=40mm]{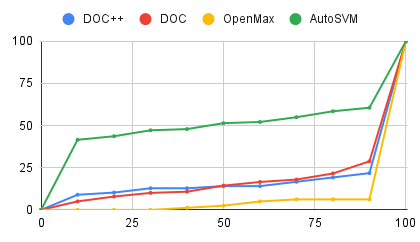} &   \includegraphics[width=40mm]{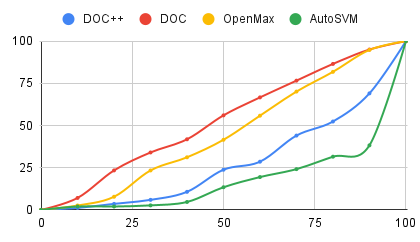} \\
		Port Scan & Botnet \\[6pt]
		\includegraphics[width=40mm]{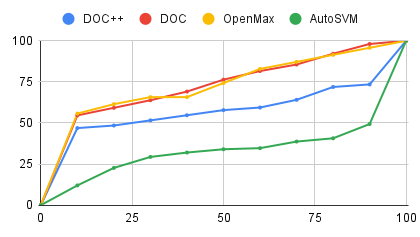} &   \includegraphics[width=40mm]{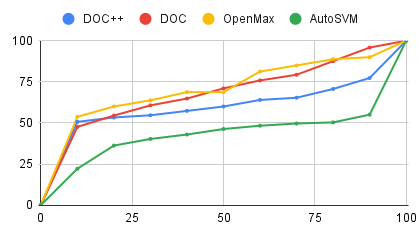} \\
		Web Attack - BruteForce & Web Attack - XSS \\[6pt]
		\includegraphics[width=40mm]{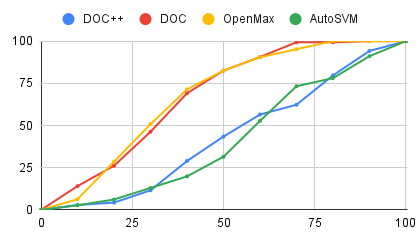} &\\
		DDoS & \\[6pt]
	\end{tabular}
	\caption{The CDF diagrams of accuracy for each attack label evaluated as the unknown (\ie, novelty) set. The vertical axes represent the accuracy of models and the horizontal axes show the percentage of experiments.}
	\label{fig: cdf_unkown_classification}
\end{figure}

In this experiment, we show the accuracy of models when trying to reject samples as unknown (novelty). The results are provided in Table \ref{tab: avg_unkown_classification} and Figure \ref{fig: cdf_unkown_classification}.
Table \ref{tab: avg_unkown_classification} shows the average accuracy of each label in several experiments when they are selected as an unknown label in the test phase.

In Figure \ref{fig: cdf_unkown_classification}, the cumulative distribution function (CDF) diagram of each attack label is presented. The CDF helps to inform the actual accuracy after removing the low accuracy outliers caused by some specific sets of labels in the training process. In other words,  if the novelty label is similar to a pre-known attack in the model (\ie, the training label set), there is a good probability that it is classified as the related attack type. On the other hand, if the novelty attack is entirely distinct from the existing classes, it will have a high-performance result.  Consequently, the CDF alongside the similarity tables (like Tables \ref{tab: similar classification} and \ref{tab:similarity_clustering_DOC++}) can give a comprehensive view of the model's capabilities (more discussion is presented in Section \ref{sec:discussion}).

In our evaluation, the best candidate implementation is the model with a lower area under the curve (AUC). That is, its most growth is in the high values for the model accuracy. It means that among all the experiments, the model has fewer experiments with low accuracy, and in most of the training/testing combinations, the model has high accuracy.

\begin{table}[!htp]\centering
	\caption{Average Accuracy on Open Set Using CNN.}\label{tab: avg_unkown_classification}
	\normalsize
	\centering
	\resizebox{\linewidth}{!}{%
		\begin{tabular}{lccccccc}
			\hline
			\parbox[c][0.5cm][c]{2.5cm}{\raggedright Labels} &DOC++ &DOC++\_NN &DOC &DOC\_NN &OpenMax &OpenMax\_NN &AutoSVM\_NN \\
			\hline
			\parbox[c][0.5cm][c]{3cm}{\raggedright \cellcolor[HTML]{b6d7a8}Botnet} &\cellcolor[HTML]{b6d7a8}63.14 &\cellcolor[HTML]{b6d7a8}71.82 &\cellcolor[HTML]{b6d7a8}45.19 &\cellcolor[HTML]{b6d7a8}45.87 &\cellcolor[HTML]{b6d7a8}48.63 &\cellcolor[HTML]{b6d7a8}54.53 &\cellcolor[HTML]{b6d7a8}83.02 \\
			\parbox[c][0.5cm][c]{3cm}{\raggedright \cellcolor[HTML]{b6d7a8}DDoS} &\cellcolor[HTML]{b6d7a8}47.64 &\cellcolor[HTML]{b6d7a8}56.56 &\cellcolor[HTML]{b6d7a8}32.03 &\cellcolor[HTML]{b6d7a8}32.38 &\cellcolor[HTML]{b6d7a8}31.21 &\cellcolor[HTML]{b6d7a8}32.45 &\cellcolor[HTML]{b6d7a8}58.03 \\
			\parbox[c][0.5cm][c]{3cm}{\raggedright \cellcolor[HTML]{ea9999}Port Scan} &\cellcolor[HTML]{ea9999}87.61 &\cellcolor[HTML]{ea9999}85.55 &\cellcolor[HTML]{ea9999}84.45 &\cellcolor[HTML]{ea9999}84.74 &\cellcolor[HTML]{ea9999}98.48 &\cellcolor[HTML]{ea9999}96.72 &\cellcolor[HTML]{ea9999}49.18 \\
			\parbox[c][0.5cm][c]{3cm}{\raggedright Benign Wednesday} &50.01 &N/A &35.01 &N/A &30.67 &N/A &N/A \\
			\parbox[c][0.5cm][c]{3cm}{\raggedright \cellcolor[HTML]{b6d7a8}DoS GoldenEye} &\cellcolor[HTML]{b6d7a8}54.76 &\cellcolor[HTML]{b6d7a8}50.26 &\cellcolor[HTML]{b6d7a8}36.05 &\cellcolor[HTML]{b6d7a8}27.58 &\cellcolor[HTML]{b6d7a8}40.68 &\cellcolor[HTML]{b6d7a8}30.3 &\cellcolor[HTML]{b6d7a8}47.06 \\
			\parbox[c][0.5cm][c]{3cm}{\raggedright \cellcolor[HTML]{b6d7a8}DoS Hulk} &\cellcolor[HTML]{b6d7a8}49.01 &\cellcolor[HTML]{b6d7a8}47.08 &\cellcolor[HTML]{b6d7a8}30.76 &\cellcolor[HTML]{b6d7a8}22.75 &\cellcolor[HTML]{b6d7a8}37.47 &\cellcolor[HTML]{b6d7a8}25.74 &\cellcolor[HTML]{b6d7a8}34.34 \\
			\parbox[c][0.5cm][c]{3cm}{\raggedright \cellcolor[HTML]{a2c4c9}DoS SlowHttp} &\cellcolor[HTML]{a2c4c9}63.09 &\cellcolor[HTML]{a2c4c9}69.13 &\cellcolor[HTML]{a2c4c9}50.69 &\cellcolor[HTML]{a2c4c9}48.12 &\cellcolor[HTML]{a2c4c9}60.63 &\cellcolor[HTML]{a2c4c9}56.02 &\cellcolor[HTML]{a2c4c9}76.74 \\
			\parbox[c][0.5cm][c]{3cm}{\raggedright \cellcolor[HTML]{a2c4c9}DoS SlowLoris} &\cellcolor[HTML]{a2c4c9}60.16 &\cellcolor[HTML]{a2c4c9}54.46 &\cellcolor[HTML]{a2c4c9}48.02 &\cellcolor[HTML]{a2c4c9}40.9 &\cellcolor[HTML]{a2c4c9}55.19 &\cellcolor[HTML]{a2c4c9}53.91 &\cellcolor[HTML]{a2c4c9}48.71 \\
			\parbox[c][0.5cm][c]{3cm}{\raggedright \cellcolor[HTML]{a2c4c9}FTP-Patator} &\cellcolor[HTML]{a2c4c9}50.04 &\cellcolor[HTML]{a2c4c9}52.19 &\cellcolor[HTML]{a2c4c9}39.46 &\cellcolor[HTML]{a2c4c9}37.37 &\cellcolor[HTML]{a2c4c9}31.13 &\cellcolor[HTML]{a2c4c9}36.46 &\cellcolor[HTML]{a2c4c9}65.05 \\
			\parbox[c][0.5cm][c]{3cm}{\raggedright \cellcolor[HTML]{b6d7a8}SSH-Patator} &\cellcolor[HTML]{b6d7a8}51.34 &\cellcolor[HTML]{b6d7a8}61.92 &\cellcolor[HTML]{b6d7a8}31.25 &\cellcolor[HTML]{b6d7a8}48.66 &\cellcolor[HTML]{b6d7a8}23.08 &\cellcolor[HTML]{b6d7a8}34.24 &\cellcolor[HTML]{b6d7a8}71.45 \\
			\parbox[c][0.5cm][c]{3cm}{\raggedright Benign Friday} &50.46 &N/A &39.39 &N/A &47.5 &N/A &N/A \\
			\parbox[c][0.5cm][c]{3cm}{\raggedright \cellcolor[HTML]{ffe599}Web BruteForce} &\cellcolor[HTML]{ffe599}40.44 &\cellcolor[HTML]{ffe599}42.24 &\cellcolor[HTML]{ffe599}22.82 &\cellcolor[HTML]{ffe599}25.8 &\cellcolor[HTML]{ffe599}29.38 &\cellcolor[HTML]{ffe599}26.09 &\cellcolor[HTML]{ffe599}66.5 \\
			\parbox[c][0.5cm][c]{3cm}{\raggedright \cellcolor[HTML]{ffe599}Web XSS} &\cellcolor[HTML]{ffe599}37.33 &\cellcolor[HTML]{ffe599}38.92 &\cellcolor[HTML]{ffe599}32.3 &\cellcolor[HTML]{ffe599}29.61 &\cellcolor[HTML]{ffe599}29.59 &\cellcolor[HTML]{ffe599}27.16 &\cellcolor[HTML]{ffe599}56.68 \\
			\hline
		\end{tabular}
	}
\end{table}

\begin{table}[!htp]\centering
	\caption{Average Accuracy on Open Set Using LSTM Cells.}
	\label{tab: LSTM Unknown}
	\scriptsize
	\begin{tabular}{lcc}
		\hline
		\parbox[c][0.25cm][c]{2.5cm}{\raggedright Labels}  &DOC++ & DOC++\_NN \\
		\hline
		\parbox[c][0.25cm][c]{3cm}{\raggedright \cellcolor[HTML]{b6d7a8}Botnet} &\cellcolor[HTML]{b6d7a8}49.01 &\cellcolor[HTML]{b6d7a8}55.41 \\
		\parbox[c][0.25cm][c]{3cm}{\raggedright \cellcolor[HTML]{b6d7a8}DDoS} &\cellcolor[HTML]{b6d7a8}48.13 &\cellcolor[HTML]{b6d7a8}38.45 \\
		\parbox[c][0.25cm][c]{3cm}{\raggedright \cellcolor[HTML]{ea9999}Port Scan} &\cellcolor[HTML]{ea9999}66.05 &\cellcolor[HTML]{ea9999}61.78 \\
		\parbox[c][0.25cm][c]{3cm}{\raggedright Benign Wednesday}  &28.29 & N/A \\
		\parbox[c][0.25cm][c]{3cm}{\raggedright \cellcolor[HTML]{b6d7a8}DoS GoldenEye} &\cellcolor[HTML]{b6d7a8}57.35 &\cellcolor[HTML]{b6d7a8}49.12 \\
		\parbox[c][0.25cm][c]{3cm}{\raggedright \cellcolor[HTML]{b6d7a8}DoS Hulk} &\cellcolor[HTML]{b6d7a8}44.9 &\cellcolor[HTML]{b6d7a8}41.82 \\
		\parbox[c][0.25cm][c]{3cm}{\raggedright \cellcolor[HTML]{a2c4c9}DoS SlowHttp} &\cellcolor[HTML]{a2c4c9}40.41 &\cellcolor[HTML]{a2c4c9}38.83 \\
		\parbox[c][0.25cm][c]{3cm}{\raggedright \cellcolor[HTML]{a2c4c9}DoS SlowLoris} &\cellcolor[HTML]{a2c4c9}52.77 &\cellcolor[HTML]{a2c4c9}41.74 \\
		\parbox[c][0.25cm][c]{3cm}{\raggedright \cellcolor[HTML]{a2c4c9}FTP-Patator} &\cellcolor[HTML]{a2c4c9}68.18 &\cellcolor[HTML]{a2c4c9}56.1 \\
		\parbox[c][0.25cm][c]{3cm}{\raggedright \cellcolor[HTML]{b6d7a8}SSH-Patator} &\cellcolor[HTML]{b6d7a8}45.39 &\cellcolor[HTML]{b6d7a8}52.64 \\
		\parbox[c][0.25cm][c]{3cm}{\raggedright Benign Friday} &28.77 & N/A \\
		\parbox[c][0.25cm][c]{3cm}{\raggedright \cellcolor[HTML]{ffe599}Web BruteForce} &\cellcolor[HTML]{ffe599}42.31 &\cellcolor[HTML]{ffe599}43.21 \\
		\parbox[c][0.25cm][c]{3cm}{\raggedright \cellcolor[HTML]{ffe599}Web XSS} &\cellcolor[HTML]{ffe599}37.1 &\cellcolor[HTML]{ffe599}39.45 \\
		\hline
	\end{tabular}
\end{table}

\subsection{Clustering and Post Train}
In this section, we present the results of the evaluation of the second phase of the framework. This phase deals with the clustering and the improvement achievable during the post-training process.
The metrics which we used in our experiments are \textbf{Completeness} and \textbf{Homogeneity}. Completeness measures the percentage of the same labels that are grouped together by the clustering algorithm. In contrast, a clustering result satisfies homogeneity if all of its clusters contain only data points that are members of a single class \cite{sckitLearnWebsite}.

The scores are calculated for both the training phase and the test phase separately. The completeness results are available in Tables \ref{tab: completeness-train} and \ref{tab: completeness-test}, respectively. Besides, the results of the homogeneity metric are reported in Tables \ref{tab:homogeneity-train} and \ref{tab:homogeneity-test}.
The results are calculated as the average percentage of different clusters' members of that specific label.
The results present the effectiveness of the post-training phase in improving both the completeness and homogeneity scores.

\begin{table}[!htp]\centering
	\caption{Completeness on Training Data.}
	\label{tab: completeness-train}
	\normalsize
	\centering
	\resizebox{\linewidth}{!}{%
		\begin{tabular}{|l|lr|lr|lr|}
			\hline
			\parbox[c][0.5cm][c]{2.5cm}{\raggedright Label} &DOC++ &DOC++\_Post &DOC &DOC\_Post &OpenMax &OpenMax\_Post \\
			\hline
			\parbox[c][0.5cm][c]{2.5cm}{\raggedright Botnet} &99.82 &99.99 &99.91 &100 &99.74 &100 \\
			\parbox[c][0.5cm][c]{2.5cm}{\raggedright DDoS} &99.48 &100 &99.27 &99.91 &99.84 &99.99 \\
			\parbox[c][0.5cm][c]{2.5cm}{\raggedright Port Scan} &99.98 &99.99 &99.93 &99.98 &99.92 &100 \\
			\parbox[c][0.5cm][c]{2.5cm}{\raggedright DoS GoldenEye} &99.1 &99.99 &99.48 &99.92 &99.35 &99.99 \\
			\parbox[c][0.5cm][c]{2.5cm}{\raggedright DoS Hulk} &99.43 &99.8 &99.23 &99.81 &99.64 &99.96 \\
			\parbox[c][0.5cm][c]{2.5cm}{\raggedright DoS SlowHttp} &99.59 &99.73 &99.68 &99.97 &99.29 &99.56 \\
			\parbox[c][0.5cm][c]{2.5cm}{\raggedright DoS SlowLoris} &99.38 &100 &99.63 &99.99 &99.6 &99.99 \\
			\parbox[c][0.5cm][c]{2.5cm}{\raggedright FTP-Patator} &100 &100 &100 &100 &99.92 &99.83 \\
			\parbox[c][0.5cm][c]{2.5cm}{\raggedright SSH-Patator} &99.86 &99.99 &99.96 &99.99 &99.98 &100 \\
			\parbox[c][0.5cm][c]{2.5cm}{\raggedright Web BruteForce} &99.65 &100 &99.98 &100 &94.99 &99.96 \\
			\parbox[c][0.5cm][c]{2.5cm}{\raggedright Web XSS} &99.77 &99.86 &99.79 &99.86 &99.75 &99.8 \\
			\hline
		\end{tabular}
	}
\end{table}

\begin{table}[!htp]\centering
	\caption{Completeness on Test Data.}
	\label{tab: completeness-test}
	\normalsize
	\centering
	\resizebox{\linewidth}{!}{%
		\begin{tabular}{|l|lr|lr|lr|}
			\hline
			\parbox[c][0.5cm][c]{2.5cm}{\raggedright Label}  &DOC++ &DOC++\_Post &DOC &DOC\_Post &OpenMax &OpenMax\_Post \\
			\hline
			\parbox[c][0.5cm][c]{3cm}{\raggedright Botnet} &67.44 &58.39 &54.6 &63.65 &67.18 &76.08 \\
			\parbox[c][0.5cm][c]{3cm}{\raggedright DDoS} &60.03 &61.47 &57.02 &66.08 &63.83 &67.53 \\
			\parbox[c][0.5cm][c]{3cm}{\raggedright Port Scan} &48.94 &92.33 &60.15 &61.58 &83.23 &86.26 \\
			\parbox[c][0.5cm][c]{3cm}{\raggedright DoS GoldenEye} &79.86 &79.63 &69.01 &69.6 &54.82 &56.06 \\
			\parbox[c][0.5cm][c]{3cm}{\raggedright DoS Hulk} &81.26 &72.46 &64.43 &58.33 &52.73 &51.1 \\
			\parbox[c][0.5cm][c]{3cm}{\raggedright DoS SlowHttp} &58.77 &63.06 &42.53 &54.09 &53.43 &61.18 \\
			\parbox[c][0.5cm][c]{3cm}{\raggedright DoS SlowLoris} &49.99 &74.97 &62.64 &63.61 &46.27 &55.7 \\
			\parbox[c][0.5cm][c]{3cm}{\raggedright FTP-Patator} &58.59 &80.51 &69.6 &86.65 &80.95 &92.28 \\
			\parbox[c][0.5cm][c]{3cm}{\raggedright SSH-Patator} &66.07 &70.39 &88.25 &94.32 &73.96 &89.3 \\
			\parbox[c][0.5cm][c]{3cm}{\raggedright Web BruteForce} &85.98 &92.19 &88.07 &67.42 &78.66 &86.09 \\
			\parbox[c][0.5cm][c]{3cm}{\raggedright Web XSS} &91.24 &71.59 &83.66 &61.8 &84.4 &92.14 \\
			\hline
		\end{tabular}
	}
\end{table}

\begin{table}[!htp]\centering
	\caption{Homogeneity on Training Data.}
	\label{tab:homogeneity-train}
	\normalsize
	\centering
	\resizebox{\linewidth}{!}{%
		\begin{tabular}{|l|lr|lr|lr|}
			\hline
			\parbox[c][0.5cm][c]{2.5cm}{\raggedright Label}  &DOC++ &DOC++\_Post &DOC &DOC\_Post &OpenMax &OpenMax\_Post \\
			\hline
			\parbox[c][0.5cm][c]{3cm}{\raggedright Botnet} &99.63 &99.98 &99.43 &99.9 &99.68 &99.96 \\
			\parbox[c][0.5cm][c]{3cm}{\raggedright DDoS} &99.79 &100 &99.88 &99.98 &99.9 &100 \\
			\parbox[c][0.5cm][c]{3cm}{\raggedright Port Scan} &99.48 &99.98 &99.68 &99.94 &99.71 &99.97 \\
			\parbox[c][0.5cm][c]{3cm}{\raggedright DoS GoldenEye} &99.08 &99.79 &99.15 &99.82 &98.71 &99.9 \\
			\parbox[c][0.5cm][c]{3cm}{\raggedright DoS Hulk} &99.84 &99.98 &99.87 &99.99 &99.82 &100 \\
			\parbox[c][0.5cm][c]{3cm}{\raggedright DoS SlowHttp} &99.8 &99.93 &99.72 &99.9 &99.73 &99.98 \\
			\parbox[c][0.5cm][c]{3cm}{\raggedright DoS SlowLoris} &99.64 &99.77 &99.67 &99.97 &99.71 &99.83 \\
			\parbox[c][0.5cm][c]{3cm}{\raggedright FTP-Patator} &99.96 &100 &100 &100 &99.96 &99.98 \\
			\parbox[c][0.5cm][c]{3cm}{\raggedright SSH-Patator} &99.95 &100 &99.92 &100 &99.97 &99.99 \\
			\parbox[c][0.5cm][c]{3cm}{\raggedright Web BruteForce} &76.75 &76.92 &53.68 &53.91 &72.8 &74.15 \\
			\parbox[c][0.5cm][c]{3cm}{\raggedright Web XSS} &96.05 &96.18 &95.98 &96.19 &75.71 &74.68 \\
			\hline
		\end{tabular}
	}
\end{table}

\begin{table}[!htp]\centering
	\caption{Homogeneity on Test Data.}
	\label{tab:homogeneity-test}
	\normalsize
	\centering
	\resizebox{\linewidth}{!}{%
		\begin{tabular}{|l|lr|lr|lr|}
			\hline
			\parbox[c][0.5cm][c]{2.5cm}{\raggedright Label}  &DOC++ &DOC++\_Post &DOC &DOC\_Post &OpenMax &OpenMax\_Post \\
			\hline
			\parbox[c][0.5cm][c]{3cm}{\raggedright Botnet} &33.7 &35.12 &21.6 &22.77 &20.7 &38.05 \\
			\parbox[c][0.5cm][c]{3cm}{\raggedright DDoS} &28.71 &31.27 &32.19 &33.85 &31.92 &26.82 \\
			\parbox[c][0.5cm][c]{3cm}{\raggedright Port Scan} &55.34 &59.79 &41.17 &37.98 &48.28 &35.46 \\
			\parbox[c][0.5cm][c]{3cm}{\raggedright DoS GoldenEye} &36.95 &39.03 &39.76 &41.34 &21.39 &20.87 \\
			\parbox[c][0.5cm][c]{3cm}{\raggedright DoS Hulk} &36.53 &36 &36.64 &40.85 &30.03 &24.44 \\
			\parbox[c][0.5cm][c]{3cm}{\raggedright DoS SlowHttp} &24.77 &25.12 &28.34 &36.04 &35.71 &31.31 \\
			\parbox[c][0.5cm][c]{3cm}{\raggedright DoS SlowLoris} &48.43 &49.51 &45.26 &49.61 &53.87 &51.36 \\
			\parbox[c][0.5cm][c]{3cm}{\raggedright FTP-Patator} &47.8 &45.75 &44.39 &40.97 &59.28 &53.41 \\
			\parbox[c][0.5cm][c]{3cm}{\raggedright SSH-Patator} &48.56 &30.04 &35.8 &49.44 &53.28 &41.1 \\
			\parbox[c][0.5cm][c]{3cm}{\raggedright Web BruteForce} &38.45 &35.27 &45.72 &41.87 &25.28 &24.03 \\
			\parbox[c][0.5cm][c]{3cm}{\raggedright Web XSS} &39.84 &45.24 &25.02 &26.73 &25.53 &24.32 \\
			\hline
		\end{tabular}
	}
\end{table}

\subsection{Classification of Unknown Labels After Post Train}
After the post-training process, we need to make sure that the process has not affected the previous results (stated in Table \ref{tab: avg_unkown_classification}) negatively. So we have performed the same experiment on the new model after the post-training process (shown in Table \ref{tab: avg_post}).

\begin{table}[!htp]\centering
	\caption{Average Accuracy On Open Set after the Post-Training Process.}\label{tab: avg_post}
	\normalsize
	\centering
	\resizebox{0.7\linewidth}{!}{%
	\begin{tabular}{lccc}
		\hline
		Labels &DOC++ & DOC &OpenMax\\
		\hline
		\cellcolor[HTML]{b6d7a8}Botnet &\cellcolor[HTML]{b6d7a8}66.15 &\cellcolor[HTML]{b6d7a8}46.38 &\cellcolor[HTML]{b6d7a8}55.41 \\
		\cellcolor[HTML]{b6d7a8}DDoS &\cellcolor[HTML]{b6d7a8}51.47 &\cellcolor[HTML]{b6d7a8}30.94 &\cellcolor[HTML]{b6d7a8}28.55 \\
		\cellcolor[HTML]{ea9999}Port Scan &\cellcolor[HTML]{ea9999}81.69 &\cellcolor[HTML]{ea9999}78.86 &\cellcolor[HTML]{ea9999}95.66 \\
		\cellcolor[HTML]{b6d7a8}DoS GoldenEye &\cellcolor[HTML]{b6d7a8}47.22 &\cellcolor[HTML]{b6d7a8}26.11 &\cellcolor[HTML]{b6d7a8}27.51 \\
		\cellcolor[HTML]{b6d7a8}DoS Hulk &\cellcolor[HTML]{b6d7a8}43.03 &\cellcolor[HTML]{b6d7a8}23.26 &\cellcolor[HTML]{b6d7a8}25.13 \\
		\cellcolor[HTML]{a2c4c9}DoS SlowHttp &\cellcolor[HTML]{a2c4c9}63.77 &\cellcolor[HTML]{a2c4c9}44.73 &\cellcolor[HTML]{a2c4c9}59.05\\
		\cellcolor[HTML]{a2c4c9}DoS SlowLoris &\cellcolor[HTML]{a2c4c9}55.52 &\cellcolor[HTML]{a2c4c9}38.51 &\cellcolor[HTML]{a2c4c9}50.5 \\
		\cellcolor[HTML]{a2c4c9}FTP-Patator &\cellcolor[HTML]{a2c4c9}49.89 &\cellcolor[HTML]{a2c4c9}35.74 &\cellcolor[HTML]{a2c4c9}38.68 \\
		\cellcolor[HTML]{b6d7a8}SSH-Patator &\cellcolor[HTML]{b6d7a8}63.22 &\cellcolor[HTML]{b6d7a8}42.91 &\cellcolor[HTML]{b6d7a8}30.12 \\
		\cellcolor[HTML]{ffe599}Web BruteForce &\cellcolor[HTML]{ffe599}38.45 &\cellcolor[HTML]{ffe599}27.98 &\cellcolor[HTML]{ffe599}34.13 \\
		\cellcolor[HTML]{ffe599}Web XSS &\cellcolor[HTML]{ffe599}39.27 &\cellcolor[HTML]{ffe599}25.69 &\cellcolor[HTML]{ffe599}35.23 \\
		\hline
	\end{tabular}
	}
\end{table}

\subsection{Evaluation over the CSE-CIC-IDS2018 Dataset}
According to the results of the previous subsections, DOC++ and AutoSVM present the best results among implemented methods for the open set module. Table \ref{tab: model comparison} reports the required resources to train and evaluate 1000 flow instances. The evaluation's hardware is the NVIDIA GeForce GTX 1080 GPU. It is noticeable that the AutoSVM model consists of two train phases (auto-encoder and SVMs). Consequently, it needed more time and resources for the training phase in comparison with DOC++. As a result, we have selected DOC++ as our final implementation choice. To validate the algorithm's effectiveness in other datasets, the evaluation of DOC++ over the CSE-CIC-IDS2018 dataset is presented in Table~\ref{tab: 2018}.

\begin{table}[!htp]\centering
	\caption{Resource and Time Consumption of Different Methods.}
	\label{tab: model comparison}
	\scriptsize
	\begin{tabular}{|l|c|c|c|}
		\hline
		Method & Memory &Train Time &Validation Time \\
		\hline
		DOC, DOC++ &1.5Gb &30s &20s \\
		OpenMax &1.5Gb &45s &25s \\
		AutoSVM &4.5Gb &35s+36s &37s \\
		\hline
	\end{tabular}
\end{table}

\begin{table}[!htp]\centering
	\caption{Accuracy Percentage of DOC++ over the  CSE-CIC-IDS2018 Dataset.}
	\label{tab: 2018}
	\normalsize
	\centering
	\resizebox{0.8\linewidth}{!}{%
	\begin{tabular}{|l|c|c|}
		\hline
		Labels &DOC++ Known &DOC++ Unknown \\
		\hline
		SQL injection &82.34 &35.47 \\
		Infilteration &85.89 &48.43 \\
		DoS GoldenEye &93.17 &54.74 \\
		Benign &88.05 &48.79 \\
		DoS Slowloris &92.17 &57.26 \\
		DDoS &90.04 &72.16 \\
		FTP-Patator &92.41 &41.13 \\
		SSH-Patator &92.37 &42.45 \\
		Botnet &92.03 &53 \\
		Web Attacks &88.93 &40.06 \\
		\hline
	\end{tabular}
}
\end{table}


\begin{table}[!htp]\centering
	\caption{Similarity of Unknown Labels Based on the Percentage of Accepted and Misclassified Experiments.}
	\label{tab: similar classification}
	\normalsize
	\centering
	\resizebox{\linewidth}{!}{%
		\begin{tabular}{|c|lr|lr|lr|}
			\hline
			\parbox[c][0.5cm][c]{0.5cm}{} & \multicolumn{2}{c|}{Accepted Experiments}  & \multicolumn{4}{c|}{Misclassified Experiments} \\
			\hline
			\multirow{15}{*}{\begin{sideways} OpenMax \end{sideways}} & \parbox[c][0.5cm][c]{2.3cm}{\raggedright Label} &\parbox[c][0.5cm][c]{1.4cm}{\raggedleft Percentage} &\parbox[c][0.5cm][c]{2.5cm}{\raggedright 1st Similar Label} &\parbox[c][0.5cm][c]{1.4cm}{\raggedleft Percentage} &\parbox[c][0.5cm][c]{2.6cm}{\raggedright 2nd Similar Label} &\parbox[c][0.5cm][c]{1.4cm}{\raggedleft Percentage} \\
			\cline{2-7}
			& \parbox[c][0.6cm][c]{3cm}{\raggedright Web BruteForce} &26.17 &\parbox[c][0.5cm][c]{3cm}{\raggedright Web XSS} &38.93 &\parbox[c][0.5cm][c]{3cm}{\raggedright Botnet} &18.12 \\
			& \parbox[c][0.5cm][c]{3cm}{\raggedright Web XSS} &26.11 &\parbox[c][0.5cm][c]{3cm}{\raggedright Web BruteForce} &41.4 & - & - \\
			\cline{2-7}
			& \parbox[c][0.6cm][c]{3cm}{\raggedright Port Scan} &94.04 &\parbox[c][0.5cm][c]{3cm}{\raggedright DoS SlowLoris} &5.96 & - & - \\
			\cline{2-7}
			& \parbox[c][0.6cm][c]{3cm}{\raggedright Botnet} &46.67 &\parbox[c][0.5cm][c]{3cm}{\raggedright DoS GoldenEye} &16.3 &\parbox[c][0.5cm][c]{3cm}{\raggedright DoS Hulk} &16.3 \\
			& \parbox[c][0.5cm][c]{3cm}{\raggedright DDoS} &19.59 &\parbox[c][0.5cm][c]{3cm}{\raggedright DoS GoldenEye} &33.11 &\parbox[c][0.5cm][c]{3cm}{\raggedright DoS Hulk} &25 \\
			& \parbox[c][0.5cm][c]{3cm}{\raggedright DoS GoldenEye} &10.27 &\parbox[c][0.5cm][c]{3cm}{\raggedright DoS Hulk} &40.41 &\parbox[c][0.5cm][c]{3cm}{\raggedright DDoS} &25.34 \\
			& \parbox[c][0.5cm][c]{3cm}{\raggedright DoS Hulk} &6.58 &\parbox[c][0.5cm][c]{3cm}{\raggedright DoS GoldenEye} &38.82 &\parbox[c][0.5cm][c]{3cm}{\raggedright DDoS} &26.97 \\
			& \parbox[c][0.5cm][c]{3cm}{\raggedright SSH-Patator} &19.21 &\parbox[c][0.5cm][c]{3cm}{\raggedright DDoS} &30.46 &\parbox[c][0.5cm][c]{3cm}{\raggedright Botnet} &24.5 \\
			\cline{2-7}
			& \parbox[c][0.6cm][c]{3cm}{\raggedright DoS SlowHttp} &61.07 &\parbox[c][0.5cm][c]{3cm}{\raggedright Web XSS} &9.4 &\parbox[c][0.5cm][c]{3cm}{\raggedright Web BruteForce} &8.72 \\
			& \parbox[c][0.5cm][c]{3cm}{\raggedright DoS SlowLoris} &31.94 &\parbox[c][0.5cm][c]{3cm}{\raggedright DoS SlowHttp} &27.08 &\parbox[c][0.5cm][c]{3cm}{\raggedright DoS GoldenEye} &22.92 \\
			& \parbox[c][0.5cm][c]{3cm}{\raggedright FTP-Patator} &23.57 &\parbox[c][0.5cm][c]{3cm}{\raggedright DoS SlowLoris} &32.86 &\parbox[c][0.5cm][c]{3cm}{\raggedright DoS GoldenEye} &16.43 \\
			\hline

			\hline
			\hline
			\multirow{13}{*}{\begin{sideways} DOC \end{sideways}} & \parbox[c][0.6cm][c]{3cm}{\raggedright Web BruteForce} &19.11 &\parbox[c][0.5cm][c]{3cm}{\raggedright Web XSS} &42.04 &\parbox[c][0.5cm][c]{3cm}{\raggedright Botnet} &19.75 \\
			& \parbox[c][0.5cm][c]{3cm}{\raggedright Web XSS} &23.45 &\parbox[c][0.5cm][c]{3cm}{\raggedright Web BruteForce} &35.17 &\parbox[c][0.5cm][c]{3cm}{\raggedright Botnet} &24.14 \\
			\cline{2-7}
			& \parbox[c][0.6cm][c]{3cm}{\raggedright Port Scan} &80.85 &\parbox[c][0.5cm][c]{3cm}{\raggedright DoS SlowLoris} &13.48 & - & - \\
			\cline{2-7}
			& \parbox[c][0.6cm][c]{3cm}{\raggedright Botnet} &35.66 &\parbox[c][0.5cm][c]{3cm}{\raggedright DoS GoldenEye} &20.98 &\parbox[c][0.5cm][c]{3cm}{\raggedright Web BruteForce} &18.18 \\
			& \parbox[c][0.5cm][c]{3cm}{\raggedright DDoS} &20.67 &\parbox[c][0.5cm][c]{3cm}{\raggedright DoS Hulk} &29.33 &\parbox[c][0.5cm][c]{3cm}{\raggedright DoS GoldenEye} &28.67 \\
			& \parbox[c][0.5cm][c]{3cm}{\raggedright DoS GoldenEye} &8.13 &\parbox[c][0.5cm][c]{3cm}{\raggedright DoS Hulk} &36.88 &\parbox[c][0.5cm][c]{3cm}{\raggedright DDoS} &26.88 \\
			& \parbox[c][0.5cm][c]{3cm}{\raggedright DoS Hulk} &5.84 &\parbox[c][0.5cm][c]{3cm}{\raggedright DoS GoldenEye} &40.26 &\parbox[c][0.5cm][c]{3cm}{\raggedright DDoS} &24.03 \\
			& \parbox[c][0.5cm][c]{3cm}{\raggedright SSH-Patator} &37.16 &\parbox[c][0.5cm][c]{3cm}{\raggedright DDoS} &30.41 &\parbox[c][0.5cm][c]{3cm}{\raggedright DoS SlowLoris} &8.11 \\
			\cline{2-7}
			& \parbox[c][0.6cm][c]{3cm}{\raggedright DoS SlowHttp} &46.72 &\parbox[c][0.5cm][c]{3cm}{\raggedright Web BruteForce} &14.6 &\parbox[c][0.5cm][c]{3cm}{\raggedright Botnet} &13.87 \\
			& \parbox[c][0.5cm][c]{3cm}{\raggedright DoS SlowLoris} &20.26 &\parbox[c][0.5cm][c]{3cm}{\raggedright DoS SlowHttp} &34.64 &\parbox[c][0.5cm][c]{3cm}{\raggedright Port Scan} &21.57 \\
			& \parbox[c][0.5cm][c]{3cm}{\raggedright FTP-Patator} &21.62 &\parbox[c][0.5cm][c]{3cm}{\raggedright DoS SlowLoris} &37.84 &\parbox[c][0.5cm][c]{3cm}{\raggedright DoS GoldenEye} &18.24 \\
			\hline
			
			\hline
			\hline
			\multirow{14}{*}{\begin{sideways} DOC++ \end{sideways}} &\parbox[c][0.6cm][c]{3cm}{\raggedright Web BruteForce} &35.92 &\parbox[c][0.5cm][c]{3cm}{\raggedright Web XSS} &38.73 &\parbox[c][0.5cm][c]{3cm}{\raggedright Botnet} &14.79 \\
			&\parbox[c][0.5cm][c]{3cm}{\raggedright Web XSS} &34.69 &\parbox[c][0.5cm][c]{3cm}{\raggedright Web BruteForce} &38.78 &\parbox[c][0.5cm][c]{3cm}{\raggedright Botnet} &14.29 \\
			\cline{2-7}
			&\parbox[c][0.6cm][c]{3cm}{\raggedright Port Scan} &80.69 &\parbox[c][0.5cm][c]{3cm}{\raggedright DoS SlowLoris} &15.86 &\parbox[c][0.5cm][c]{3cm}{\raggedright Botnet} &2.76 \\
			\cline{2-7}
			& \parbox[c][0.6cm][c]{3cm}{\raggedright Botnet} &61.97 &\parbox[c][0.5cm][c]{3cm}{\raggedright DoS GoldenEye} &18.31 &\parbox[c][0.5cm][c]{3cm}{\raggedright Web BruteForce} &9.86 \\
			& \parbox[c][0.5cm][c]{3cm}{\raggedright DDoS} &41.18 &\parbox[c][0.5cm][c]{3cm}{\raggedright DoS Hulk} &26.47 &\parbox[c][0.5cm][c]{3cm}{\raggedright DoS GoldenEye} &19.12 \\
			& \parbox[c][0.5cm][c]{3cm}{\raggedright DoS GoldenEye} &31.03 &\parbox[c][0.5cm][c]{3cm}{\raggedright DoS Hulk} &38.62 &\parbox[c][0.5cm][c]{3cm}{\raggedright DDoS} &22.07 \\
			& \parbox[c][0.5cm][c]{3cm}{\raggedright DoS Hulk} &24.63 &\parbox[c][0.5cm][c]{3cm}{\raggedright DoS GoldenEye} &41.79 &\parbox[c][0.5cm][c]{3cm}{\raggedright DDoS} &21.64 \\
			& \parbox[c][0.5cm][c]{3cm}{\raggedright SSH-Patator} &49.23 &\parbox[c][0.5cm][c]{3cm}{\raggedright DDoS} &30 &\parbox[c][0.5cm][c]{3cm}{\raggedright Botnet} &13.08 \\
			\cline{2-7}
			& \parbox[c][0.6cm][c]{3cm}{\raggedright DoS SlowHttp} &65.63 &\parbox[c][0.5cm][c]{3cm}{\raggedright Botnet} &9.38 &\parbox[c][0.5cm][c]{3cm}{\raggedright DoS GoldenEye} &8.59 \\
			& \parbox[c][0.5cm][c]{3cm}{\raggedright DoS SlowLoris} &32 &\parbox[c][0.5cm][c]{3cm}{\raggedright DoS SlowHttp} &32 &\parbox[c][0.5cm][c]{3cm}{\raggedright DoS GoldenEye} &21 \\
			& \parbox[c][0.5cm][c]{3cm}{\raggedright FTP-Patator} &37.78 &\parbox[c][0.5cm][c]{3cm}{\raggedright DoS SlowLoris} &36.3 &\parbox[c][0.5cm][c]{3cm}{\raggedright DoS GoldenEye} &14.07 \\
			\hline
		\end{tabular}
	}
\end{table}

\section{Discussion}\label{sec:discussion}
\label{discussion}
In the following, the obtained results are discussed for more clarification purposes and interpretation.
\subsection{Best Deployment of the Proposed Framework}
Our experiments (Tables \ref{tab: avg_known_acc}, \ref{tab: LSTM Known}, \ref{tab: avg_unkown_classification}, and \ref{tab: LSTM Unknown}) show that when the normal traffic is included in the training labels, the accuracies of classification on both the closed set and open set problems are decreased for some labels. As a result, the best framework deployment is when a dedicated module separates the benign traffic from the abnormal one. This module can be an anomaly-based module that just traces the benign traffic. The different implementations of this module in deep learning-based IDS are topics of futher research.
\subsection{Similar Labels}
There is a pattern in misclassified samples of unknown classification. Most of the time, some specified groups of samples are misclassified interchangeably. 
Consequently, we created group labels and placed similar labels in one group. To achieve this, we used different approaches:
\subsubsection{Classification-Based Similarity}
The first method we applied to find similar labels was to check the results of unknown classification. We assumed a threshold (70 percent) for determining the misclassified experiments related to an unknown label (test label).  The experiments with higher accuracy of the determined threshold are considered ``Accepted'' experiments. If the value of the accuracy of the unknown label is less than 70 percent, the correspondent experiment is considered a misclassified experiment. These misclassified experiments usually consist of some known classes which are similar to the test label. The top known labels of different misclassified experiments for a specific unknown label make the similarity group of the target label. Table \ref{tab: similar classification} presents the similar groups and the percentage of accepted and misclassified experiments for each label in different evaluated methods.

\subsubsection{Clustering-Based Similarity}
The other method to find similar labels is to cluster the samples and check the labels closed together.
To achieve this, we used the output from the DOC++ module to run the clustering for groups of unknown labels which were not a part of the training labels. The result of this evaluation is available in Table \ref{tab:similarity_clustering_DOC++}.
In this table, we have presented each label alongside the top 3 labels it has been clustered with. The percentages show the number of experiments where the target label and the similar one are clustered together.

\begin{table*}[!ht]
		\caption{Similarity of Unknown Labels Clustering Based on the Percentage of DOC++ Experiments.}
		\label{tab:similarity_clustering_DOC++}
		\centering
		\resizebox{0.8\textwidth}{!}{%
			\begin{tabular}{|l|lr|lr|lr|lr|}
				\hline
				Label & 1st Similar Label & Percentage & 2nd Similar Label & Percentage & 3rd Similar Label & Percentage \\
				\hline
				DoS GoldenEye &DoS Hulk &15.01 &DDoS &12.16 &Botnet &8.67 \\
				DoS Hulk & DoS GoldenEye &16.03 &DDoS &12.02 &Botnet &9.54 \\
				DoS SlowHttp &DDoS &11.26 &Botnet &11.2 &DoS SlowLoris &10.52 \\
				DoS SlowLoris & DoS SlowHttp &10.74 &Port Scan &10.55 &DDoS &8.33 \\
				FTP-Patator & DDoS &10.17 &Port Scan &9.3 &DoS Hulk &9.04 \\
				SSH-Patator & DDoS &11.71 &DoS GoldenEye &8.12 &Botnet &7.96 \\
				Web BruteForce & Web XSS &17.21 &Botnet &13.72 &DDoS &7.23 \\
				Web XSS & Web BruteForce &20.14 &Botnet &14.4 &DoS SlowHttp &7 \\
				DDoS & DoS GoldenEye &10.24 &Botnet &10.13 &DoS SlowHttp &9.86 \\
				Botnet & DDoS &10.26 &DoS SlowHttp &9.93 &Web BruteForce &9.1 \\
				Port Scan & DoS SlowLoris &14.89 &FTP-Patator &9.6 &DoS SlowHttp &8.97 \\
				\hline
			\end{tabular}
		}
\end{table*}

Based on the two above approaches for finding similar labels in the CIC-IDS2017 dataset, we finally decided to create groups of labels and investigate their similarities.
The created groups are described as follows:

\begin{enumerate}
	\item DoS attacks: Botnet, DDoS, DoS GoldenEye, DoS Hulk, SSH-Patator,
	\item Web attacks: Web BruteForce, Web XSS,
	\item Portscan: Portscan,
	\item Slow-rate attacks: DoS SlowHttp, DoS SlowLoris, FTP-Patator.
\end{enumerate}

After finding similar labels using classification and clustering results, we tried to figure out whether or not the groups we created made sense.
We prepared some sample flows from each label and observed them in detail using the Wireshark tool. 

In the first group, DoS Hulk and DoS GoldenEye continuously send HTTP requests to single or multiple URLs to overwhelm web servers' resources. Other attack types of this group have similar traffic.
DDoS and Botnet attacks are based on sending similar HTTP requests from many different agents/bots. 

The second and third groups consist of web-based attacks and portscans. The main difference between portscan and other types of attacks is the usage of different ports. The fourth group consists of slow-rate attacks. These attacks send requests with high time intervals to avoid detection by the IDSes. 

It may seem strange to see SSH-Patator in the same group as the DDoS attacks. SSH-Patator and FTP-Patator are both dictionary attacks that use a list of known users and passwords to find the valid credential of a targeted server. But according to the obtained results, these two attacks are grouped differently. The main reason is that the simulated SSH-Patator sends many encrypted requests with low intervals like other high-speed DoS attacks. On the other hand, as we analyzed the PCAP traffic of the FTP-Patator attack, some non-encrypted authentication messages are sent slowly. This behavior makes the main sharing point of this attack with the other ones in the slow-rate group.

Finally, the evaluation of DOC++ with the assumption that the labels in the same similarity group are all considered as the correct result, is reported in Table \ref{tab: DOC++ Groups}. According to the proposed similarity groups, the model has high accuracy over most of the attack types. The only group that has less accurate results is the web attack category. The reason is that web attacks are mainly repeated frequently. Consequently, they will be misclassified as DDoS and Botnet attacks.

\begin{table}[!htp]\centering
	\caption{Average Accuracy on Open Set Considering Group Labels.}
	\label{tab: DOC++ Groups}
	\scriptsize
	\begin{tabular}{lc}
		\hline
		\parbox[c][0.25cm][c]{2.5cm}{\raggedright Labels}  & \parbox[c][0.25cm][c]{1cm}{DOC++} \\
		\hline
		\parbox[c][0.25cm][c]{3cm}{\raggedright \cellcolor[HTML]{b6d7a8}Botnet} &\cellcolor[HTML]{b6d7a8}99.94 \\
		\parbox[c][0.25cm][c]{3cm}{\raggedright \cellcolor[HTML]{b6d7a8}DDoS} &\cellcolor[HTML]{b6d7a8}99.94 \\
		\parbox[c][0.25cm][c]{3cm}{\raggedright \cellcolor[HTML]{ea9999}Port Scan} &\cellcolor[HTML]{ea9999}99.85 \\
		\parbox[c][0.25cm][c]{3cm}{\raggedright \cellcolor[HTML]{b6d7a8}DoS GoldenEye} &\cellcolor[HTML]{b6d7a8}100 \\
		\parbox[c][0.25cm][c]{3cm}{\raggedright \cellcolor[HTML]{b6d7a8}DoS Hulk} &\cellcolor[HTML]{b6d7a8}99.85 \\
		\parbox[c][0.25cm][c]{3cm}{\raggedright \cellcolor[HTML]{a2c4c9}DoS SlowHttp} &\cellcolor[HTML]{a2c4c9}99.76 \\
		\parbox[c][0.25cm][c]{3cm}{\raggedright \cellcolor[HTML]{a2c4c9}DoS SlowLoris} &\cellcolor[HTML]{a2c4c9}99.83 \\
		\parbox[c][0.25cm][c]{3cm}{\raggedright \cellcolor[HTML]{a2c4c9}FTP-Patator} &\cellcolor[HTML]{a2c4c9}99.8 \\
		\parbox[c][0.25cm][c]{3cm}{\raggedright \cellcolor[HTML]{b6d7a8}SSH-Patator} &\cellcolor[HTML]{b6d7a8}99.93 \\
		\parbox[c][0.25cm][c]{3cm}{\raggedright \cellcolor[HTML]{ffe599}Web BruteForce} &\cellcolor[HTML]{ffe599}37.37 \\
		\parbox[c][0.25cm][c]{3cm}{\raggedright \cellcolor[HTML]{ffe599}Web XSS} &\cellcolor[HTML]{ffe599}42.3 \\
		\bottomrule
	\end{tabular}
\end{table}


\section{Conclusion}\label{sec:conclusion}
\label{conclusion}
In this paper, we presented a framework for deep learning-based IDSes to adapt to zero-day attacks. To the best of our knowledge, this work is the first study that evaluates the deep novelty-based classifiers in the network security context. Besides, the combination of clustering algorithms and deep models is another novelty of this study to provide the sufficient samples required for the update phase of the framework. Additionally, a new version of DOC (called DOC++) is presented as an open set classifier. 

The proposed framework consists of four phases: Open set recognition, Clustering/post-training, Supervised labeling, and Updating. The evaluation of this framework has been done mainly over the CIC-IDS2017 dataset. Besides, to make the results more reliable, some evaluations were reported over the CSE-CIC-IDS2018 dataset as well. Finally, a detailed analysis of the obtained results and similar labels in novelty-based detectors and the clustering phase was presented.
\ifCLASSOPTIONcaptionsoff
  \newpage
\fi



\bibliographystyle{IEEEtran}
\bibliography{references}
\end{document}